\documentclass[aps]{revtex4}
\usepackage{graphicx}
\usepackage{epsfig}

\newcommand{\bea}{\begin{eqnarray}}
\newcommand{\eea}{\end{eqnarray}}
\def\lag{\langle}
\def\rag{\rangle}
\newcommand{\p}{{\bf p}}
\def\be{\begin{equation}}
\def\ee{\end{equation}}

\def\addtext#1{{#1}}
\def\removetext#1{{}}
\def\at#1{\addtext{#1}}

\begin{document}

\title{Second sheet $\sigma$-pole and
the  threshold enhancement of the spectral function \\
in the scalar-isoscalar meson-sector}

\author{A. Patk{\'o}s$^{a }$
\footnote{Electronic address: patkos@ludens.elte.hu}} 
\author{Zs. Sz{\'e}p$^{a }$
\footnote{Present address: Research Group for Statistical Physics
of the Hungarian Academy of Sciences,  
H-1117, Budapest, Hungary. Electronic address: szepzs@antonius.elte.hu}}
\author{ P. Sz{\'e}pfalusy$^{b,c}$
\footnote{Electronic address: psz@galahad.elte.hu}}
\affiliation{$^a$ Department of Atomic Physics,\\
$^b$ Department of Physics of Complex Systems, \\
E{\"o}tv{\"o}s University, H-1117 Budapest, Hungary\\
$^c$ Research Institute for Solid State Physics and Optics,
\hbox{Hungarian Academy of Sciences, H-1525 Budapest, Hungary}}


\begin{abstract}
The scalar-isoscalar propagator of the effective linear $\sigma$-model
of meson dynamics is investigated with the help of an expansion in the
number of the Goldstone-bosons. A generic scenario is suggested for
the temperature/density driven evolution of its pole in the second Riemann
sheet. An extended temperature range, correlated with characteristic pole 
locations, is found where the phenomenon of threshold enhancement takes
place in the corresponding spectral function.
\end{abstract}

\maketitle

\section{Introduction}

The aim of the present investigation is to relate the so-called
$\sigma$-pole in the second Riemann sheet in the complex frequency
plane and the behavior of the corresponding spectral function at finite
temperature/density. Both objects will be determined from the
scalar-isoscalar propagator. \at{The calculation suggests} 
a generic trajectory
for this pole irrespective of the nature of the thermodynamical driving
force. We investigated the question of when a well-defined resonance
characterized by a Lorentzian shaped spectral function is
present. The study
is performed in the framework of the linear $\sigma$-model, used as an 
effective field theory describing the fluctuations of the chiral order
parameter. We find the most convenient the application of an expansion in
the number of the Goldstone mesons, which is a kind of large $N$ approach to 
the physical excitation spectra of the relativistic $O(N)$ field theory in 
its broken symmetry phase.

The application of the $O(N)$ symmetric scalar field theory to the
thermal and finite baryonic density behavior of the pion-sigma system
was suggested and has been actively pursued for about 15
years, in particular by Hatsuda, Kunihiro, and
collaborators (for the latest review see \cite{hatsuda01}). The main 
physical effect proposed for the scalar-isoscalar spectral function is
its gradual enhancement near that value of the temperature/baryon density where
the phase space available for the $\sigma\rightarrow 2\pi$ decay is
squeezed to zero. 

For the theoretical consolidation of this effect, Hatsuda {\it et al.}
put forward
apparently model independent arguments for the behavior of both the
real and imaginary parts of the $\sigma$ propagator.
Using an improved version of the finite temperature/density
perturbative evaluation of the $\sigma$ self-energy \cite{chiku98}
the pole describing the $\sigma$-resonance moves
from its zero temperature/density location smoothly to the location of
the two-pion threshold. Both real and
imaginary parts of the pole location diminish monotonically. 
\at{The real part approaches the two-pion threshold faster than the
imaginary part vanishes, and the spectral function becomes
proportional in this temperature range to the inverse of the imaginary
part \cite{hatsuda99}. The maximum of the proposed threshold
enhancement occurs for that well-defined temperature/density value
when the $\sigma$-pole reaches the threshold.}

 The effects of partial symmetry restoration realized by the
 diminishing of $f_\pi$ was studied also
on the unitarized $\pi -\pi$ scattering amplitude \cite{jido01}. 
This quantity was computed and analyzed recently  with dispersive
 techniques in the framework of the chiral perturbation theory
\cite{yokokawa02}.

The approximations made by these authors may be verified soon when reliable 
spectral functions will be obtained using the non-perturbative lattice field
theoretical approach \cite{asakawa00,wetzorke01}. In the meantime
different semi-analytic approximation schemes also might shed light on the
generality of the proposed arguments. A possible scheme can be based
on an expansion in the inverse number of the Goldstone bosons.

The large $N$ expansion has been applied already some 30 years ago to
the characterization of the elementary excitations of critical
$O(N)$ symmetric lattice systems \cite{sasvari74,sasvari75}. It was
realized that it provides faithful information on the excitation
spectra of these systems in the full broken symmetry phase between
zero and the critical temperature. 

In a recent letter \cite{patkos02} we proposed its
application to the present relativistic system, since it avoids all
problems of principles showing up in other perturbative approaches.
First of all its validity does not depend on the rather strong
self-coupling of the effective $\sigma$-model. Second, its result is
not sensitive to the choice of the normalization point, that is its
change can be always compensated by an appropriate change in the value of the
couplings as required by the renormalization group. Third, it leads
automatically to the same spectra for the elementary $\sigma$-field
and for the scalar-isoscalar quadratic composite field. In other
approaches this feature is usually missing or found to be true
after non-trivial manipulations. Finally, in the chiral limit it
provides a correct critical description of the chiral symmetry restoration.

The leading order large $N$ approximation has been applied to the Goldstone 
boson scattering by Chivukula and Golden \cite{chivukula92} at zero 
temperature. They have explicitly checked that this approximation fulfills in
the scalar-isoscalar channel the unitarity condition and also satisfies
the Adler-zero condition.

The large $N$ leading order amplitude $(N-1)A(s)$ has been compared also with 
the existing phase shift data in the $I=J=0$ channel of the $\pi-\pi$
scattering by Dobado and Morales \cite{dobado95}. They have completed the
leading order amplitude by subleading terms, dictated by the requirement
of crossing symmetry: $A_{00}(s,t,u)=3A(s)+A(t)+A(u)$.
Though this is not a systematic next-to-leading order
computation the authors found a satisfactory fit to the relevant phase
shift $\delta_0^0(s)$ up
to $\sqrt{s}\approx 600 $MeV.

To our best knowledge no application of the large $N$ approach at finite 
temperature/baryon density was attempted to date. In view of the
importance of $t$- 
and $u$-channel exchange contributions to the $\pi-\pi$ scattering amplitude,
emphasized in the recent literature \cite{anisovich00}, a fully satisfactory
large $N$ treatment will also require the calculation of the next-to-leading
order contribution to the $\sigma$ propagator. The leading order
calculation presented in this paper
is a necessary intermediate step also towards this goal.

\at{The fact that for $T=0$ the real and imaginary parts of
the $\sigma$-resonance are of comparable magnitude implies that for its
finite temperature description we cannot restrict ourselves to the
immediate vicinity of the real axis. An exception is when the
$\sigma$-resonance gets close to the two-pion threshold, which is the
situation where threshold enhancement develops. For the explanation of
the detailed features of this phenomenon it is unavoidable to trace
the complete temperature driven pole trajectory. The main purpose of
the present paper is to carry out such an investigation within the
leading order of the large N expansion.}
According to our calculation to be presented in this paper,
the $\sigma$ self-energy continued
analytically into the lower halfplane leads to a pole trajectory 
(when the temperature $T$ or the baryonic density, $n_B$ is varied) whose 
real part assumes smaller values than twice the pion mass, while the
decrease of the imaginary part is not drastic. 
As a consequence of this the pole looses its meaning as a resonance and at
the same time in the spectral function the direct neighborhood of the
threshold will be emphasized. \at{When further
increasing $T/n_B$ the trajectory hits the real axis on the second
Riemann sheet and moves along it before it reaches the threshold. This
pole evolution will be demonstrated to be generic in the sense that it is
insensitive to the variation of the parameters of the theory.
One can argue that it can stay valid beyond the leading order 
large N approximation, since it is unlikely that the trajectory
would hit the threshold point directly.} 
Only in the limiting chiral symmetric case do we 
find the smooth behavior which was proposed  by Hatsuda {\it et al.}
\cite{hatsuda01}.

One should be aware of the fact that in scalar models a tachyonic pole
is always present \cite{cooper94,cooper97,boyan99}
related to the Landau-ghost phenomenon. It restricts the range of
variation of the renormalized
parameters where the model can be used in an effective sense at all.
For this reason we find in the leading large $N$ approximation that 
a parametrization accounting for the $T=0$
phenomenological data of the $\pi$ and $\sigma$ mesons can be
achieved only approximately. However, we shall argue that
the relationship between the pole trajectory and the variation of the
spectral function to be described below might not change qualitatively.
   
The presentation of the paper is aimed at a self-contained, 
technically transparent description. In section II the Schwinger-Dyson
equations for the finite temperature two-point functions of the linear
$\sigma$ model are given to leading order in $N$. Their analytical
continuation onto the second Riemann sheet in the complex frequency
variable is presented in section III. Its explicit expression was found
 by studying the  bubble diagram, describing the splitting of
the $\sigma$-field into two off-shell pions. In section IV we
analyze the temperature driven evolution of the $\sigma$
pole-trajectory.
This investigation makes use of the $T=0$ parametrization of the model as
an input, therefore one first works out the details of the physical
and unphysical poles for $T=0$. Here we fix all couplings in a way to
achieve the closest possible characteristics of the $\sigma$ to the 
particle data, and still staying by a factor of 2 to 3 below the ``energy scale'' 
of the tachyon. In the second part of the section we shall argue that the
pole-trajectory found by varying the temperature remains qualitatively
the same when the system evolves under the variation of the baryonic
density. A detailed discussion of the change in the pole trajectory pattern
with the pion mass is also presented.
The spectral function of $\sigma$ is computed in section V 
by approaching the real axis from the physical upper half-plane. 
We shall analyze also the function which arises when 
\at{the threshold factor} is divided out. It will be demonstrated that
in an extended temperature range the maximum of the spectral
function is located in the closest neighborhood of the ($T$-dependent)
position of the two-pion threshold. The extension of this interval is
very well correlated with \at{the} piece of the pole
trajectory, \at{when its real part is below the two-pion threshold.}
In section VI the conclusions of the present study are summarized.

\section
{ Leading large $N$ expression of the propagators at finite temperature}

The appropriate parametrization of the $O(N)$ symmetric Lagrangian for
 a large $N$ expansion has the following form:
\begin{equation}
L={1 \over 2}[\partial_\mu\phi^a\partial^\mu\phi^a-m^2\phi^a\phi^a]-
{\lambda\over 24N} (\phi^a)^2(\phi^b)^2+\sqrt{N}h\phi^1.
\label{Eq:Lagra}
\end{equation}
The last term explicitly breaks the $O(N)$ symmetry and introduces 
non-zero mass for pions.

In the broken symmetry phase one separates the expectation value $\Phi(T)$ 
of the field, which points along the direction $a=1$ in the internal space 
\begin{equation}
\phi^a\rightarrow (\sqrt{N}\Phi(T)+\phi^1,\phi^i).
\label{shift}
\end{equation}
\at{In the following all quantities will be computed to leading order
in the large N limit.}

The quantum fluctuations of the order parameter are divided
into a longitudinal mode, which represents the $\sigma$ meson and 
the transversal ones. The latter correspond to the Goldstone modes, the pions.
\at{Their mass, $m_G(T)$ is determined as the pole of the resummed pion
propagator, in which the tadpole contribution is calculated with the
pion propagator selfconsistently. This results in the following gap equation:}
\bea
m_G^2(T)&=&m^2+{\lambda \over 6}\Phi^2(T)+{\lambda\over 6N}
\lag(\phi^a)^2\rag
=m^2+{\lambda \over 6}\Phi^2(T)+{\lambda\over 6}
\int\frac{d^3 k}{(2\pi)^3}\frac{1}{2\omega_k}\left(1+2n(\omega_k)\right),
\label{Eq:M2}
\eea
where $n(\omega_k)=1/(\exp(\omega_k/T)-1)$ and  $\omega_k=\sqrt{k^2+m_G^2(T)}$.

The equation of state obtained from the requirement 
$\lag\phi^1\rag=0$ is as follows:
\be
 \sqrt{N}\Phi(T) \Bigl[m^2+{\lambda \over 6}\Phi^2(T)+{\lambda\over 6}
\int\frac{d^3 k}{(2\pi)^3}\frac{1}{2\omega_k}\left(1+2n(\omega_k)\right)
-{h\over\Phi(T)}\Bigr ]=0.
\label{eqstate}
\ee
Comparing this with Eq.(\ref{Eq:M2}) one can observe that for 
$\Phi(T)\neq0$  consistency requires $\displaystyle m_G^2(T)={h/\Phi(T)}$.
\at{This is precisely the Goldstone theorem in the
presence of explicit symmetry breaking.}

\begin{figure}[htbp]
\begin{center}
\includegraphics[width=16cm]{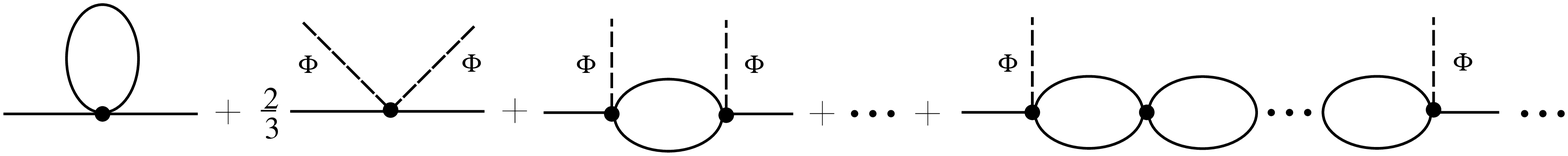}
\end{center}
\vspace*{-0.7cm}
\caption{Diagrams determining the self-energy of the $\sigma$ field 
represented to leading order in $N$. The external solid lines
correspond to the $\sigma$, while in the internal bubbles pions
propagate. The dashed line represents the expectation 
value $\Phi\equiv\Phi(T)$. \at{The vertices can be read  from 
Eq. (\ref{Eq:Lagra})
after the shift defined in Eq. (\ref{shift}) is performed.}}
\label{Fig:Graphs}
\end{figure}

At leading large $N$ order the contribution to the
longitudinal self energy is given by the sum of the
contributions corresponding to the diagrams of Fig.\ref{Fig:Graphs}, i.e. the 
bubble series. On both internal lines of a bubble exclusively the propagation 
of pion fields are taken into account to leading order in $N$. In 
each term of the bubble contribution there is a common multiplicative
vertex contribution coming from the two edges of each diagram. 
The bubble series consistently takes into account the non-zero
classical value for $\Phi(T)$, partly by the implicit dependence of 
$m_G^2(T)$, \at{due to the gap equation} and also by the two legs of the
effective  4-point vertex formed by 
the sum of the bubble-series. 

\noindent
By adding the tree level mass $m^2$ to the self-energy of the $\sigma$ field 
determined in the background $\Phi(T)$, using Eq. (\ref{eqstate}) one
finds for the 
effective $\sigma$-mass a simple expression in terms of $b(p)$, denoting the 
value of the single bubble diagram:
\begin{equation}
m_\sigma^2(p)={h\over \Phi (T)}+
\frac{\lambda}{3}\Phi^2(T)\left[1+\frac{\lambda}{6}b(p)+
\left(\frac{\lambda}{6}b(p)\right)^2+\dots
\right]
={h\over \Phi (T)}+\frac{\lambda\Phi^2(T)/3}{1-\lambda b(p)/6}.
\label{Higgs-polar}
\end{equation}

The bubble contribution with external momentum $p=(p_0,{\bf p})$ is the sum
of a zero temperature and a $T$-dependent part, 
$b^>(p)=b^>_0(p)+b^>_T(p_0,{\bf p})$.
The superscript $>$ hints at the fact that the expression of the bubble 
contribution is valid in the upper $p_0$ half-plane. The explicit expressions
of the two terms read as follows:
\bea
b_0^>(p)&=&i\int\frac{d^4k}{(2\pi)^4}\frac{1}{k^2-m_G^2(T)+i\varepsilon}
\frac{1}{(p+k)^2-m_G^2(T)+i\varepsilon},\\
b_T^>(p) &=& \int\frac{d^3{\bf q}}{(2\pi)^3}
\,\frac1{4\omega_1\omega_2}\,\left\{  (n_1+n_2)\left[
\frac{1}{p_0-\omega_1-\omega_2 + i\epsilon} - 
\frac{1}{p_0+\omega_1+\omega_2 + i\epsilon}\right]\right. \nonumber\\
&&\left.-(n_1-n_2) \left[\frac{1}{p_0-\omega_1+\omega_2 + i\epsilon} -
\frac{1} {p_0+\omega_1-\omega_2 + i\epsilon} \right]\right\},
\label{bubbleT}
\eea
where $n_i=1/(\exp(\beta\omega_i)-1)$ and $\omega_1=({\bf q}^2+m_G^2(T))^{1/2},
\omega_2=(({\bf q}+{\bf p})^2+m_G^2(T))^{1/2}$, and $\epsilon >0$.

Using cut-off regularization the zero temperature bubble contribution is:
\begin{equation}
b^>_0(p)=\frac{1}{16\pi^2}\left[\ln\frac{m_G^2(T)}{\Lambda^2}-1-
\sqrt{1-\frac{4m_G^2(T)}{p^2}}
\left(
\ln\frac{1-\sqrt{1-4m_G^2(T)/p^2}}{1+\sqrt{1-4m_G^2(T)/p^2}}+i\pi\right)
\right].
\label{Eq:unrenb0}
\end{equation}

It is common to discuss the spectral function 
for $\p=0$, when the expression of the finite temperature bubble contribution 
simplifies to:
\begin{equation}
b^>_T(p_0)=-\frac{1}{4\pi^2}\int_{m_G(T)}^\infty
\frac{dq_0}{q_0}\sqrt{q_0^2-m_G^2(T)}\left[
\frac{n(q_0)}{2q_0-p_0}+\frac{n(q_0)}{2q_0+p_0}\right],\qquad 
\textrm{Im}~p_0>0.
\label{Eq:buble_T}
\end{equation}

In view of the quadratic and logarithmic cut-off dependencies which appear
in Eqs.(\ref{Eq:M2}),(\ref{eqstate}) and (\ref{Eq:unrenb0}) a mass- and
coupling constant renormalization is necessary. It requires the introduction
of a normalization scale $M_0$. This scale should lie below the scale of the
tachyonic pole to be discussed below.
Its choice within the relevant range where the $O(N)$ model can serve for the
effective description of hadron dynamics should not affect sensitively the
physical results.

The expressions of the renormalized couplings go beyond the accuracy of
the usual one-loop relations. Actually, they ensure that a change in the
normalization scale can be compensated in all formulae below by an appropriate
change in the couplings. The following non-perturbative mass- and self-coupling
renormalizations are introduced:
\begin{equation}
{m^2\over\lambda}+{\Lambda^2\over 96\pi^2}=
{m^2_R\over\lambda_R},
\qquad {1\over\lambda} +{1\over 96\pi^2}\ln{e\Lambda^2\over
M_0^2}={1\over\lambda_R}. 
\label{renorm}
\end{equation}
In terms of the renormalized couplings the equation of state Eq.(\ref{eqstate})
can be cast into the following explicitly finite form:
\bea
&&  
{\lambda_R\over 6}{\Phi_0^2\over m_{G0}^2} 
\left({m_{G0}^4\over m_G^4(T)}-1\right)
+{\lambda_R\over 96\pi^2}\left[\left({m_G^2(T)\over m_{G0}^2}-1\right)
\ln{m_{G0}^2e\over M_0^2}+{m_G^2(T)\over m_{G0}^2}
\ln{m_G^2(T)\over m_{G0}^2}\right]
\nonumber\\
&+&
{\lambda_R T^2\over 12\pi^2 m_{G0}^2}\int_{m_G(T)/T}^\infty dy
\frac{\sqrt{y^2-m_G^2(T)/T^2}}{\exp(y)-1}={m_G^2(T)\over m_{G0}^2}-1.   
\label{Eq:ren_EoS}
\eea
Here $m_{G0}$ and $\Phi_0$ stand for the $T=0$ value of the Goldstone mass 
and of the expectation value of the $\sigma$ field, respectively.

The formal expression of the effective mass term of the $\sigma$-propagator
(\ref{Higgs-polar}) is unchanged after renormalization, just $\lambda_R$ should
replace $\lambda$ and for the $T=0$ contribution of $b(p)$ 
in Eq. (\ref{Eq:unrenb0}) the scale $M_0^2$  is put in place of $e\Lambda^2$.

The finite part of the zero temperature bubble contribution defined in this way
has different forms depending on the range of $p_0$ values:
\bea  
b_0^>(p_0)=\frac{1}{16\pi^2}\left\{
\begin{array}{l}
\displaystyle
\left[\ln\frac{m_G^2(T)}{M_0^2}+
2\sqrt{\frac{4m_G^2(T)}{p_0^2}-1}\times
\textrm{arctan}\left(\frac{4m_G^2(T)}{p_0^2}-1\right)^{-\frac{1}{2}}\right] 
,~ p_0<2m_G(T)\\
\\\displaystyle
\left[\ln\frac{m_G^2(T)}{M_0^2}-
\sqrt{1-\frac{4m_G^2(T)}{p_0^2}}
\left(
\ln\frac{1-\sqrt{1-4m_G^2(T)/p_0^2}}{1+\sqrt{1-4m_G^2(T)/p_0^2}}+i\pi\right)
\right],~ p_0>2m_G(T).
\end{array}\right.
\label{Eq:b0>R}
\eea

\section
{Analytical continuation of the propagators onto the second Riemann sheet}

The interpretation of the temperature driven variation of the spectral function
will be based on the study of the variation of the scalar-isoscalar pole mass,
that is the zero of $G_\sigma^{-1}(p_0,\p=0)=p_0^2-m_\sigma^2$ in the 
lower $p_0$ halfplane.
For this it is necessary to construct an analytical continuation of the
longitudinal propagator onto the second Riemann sheet. By 
Eq.(\ref{Higgs-polar}) it is clear that the problem is equivalent to the 
continuation of $b(p_0)$, which will be discussed in this section.

We decided to perform the  analytical continuation
in such a way that $b(p_0)$ varies continuously when 
the real axis is crossed above the two-pion threshold $p_0>2m_G(T).$ 
This implies that the bubble contribution is discontinuous across the real axis
for $-2m_G(T)<p_0<2m_G(T)$. We will actually see that there it is 
${\rm Re}~b(p_0)$ which is discontinuous.

Above the threshold on the real axis both the real and imaginary parts of the 
zero temperature bubble are continuous as one can see from the second line
of Eq. (\ref{Eq:b0>R}). In view of this, for values of $p_0$ below
the real axis (lying on the second Riemann sheet) we simply use
the expression written in the second line of Eq. (\ref{Eq:b0>R}). We denote
this continuation $b_0^<(p_0)$. (In this sense the indices ``$>$'' and
``$<$'' on $b_0$ are redundant, yet we keep them for the sake of clarity.)

We turn now to the finite temperature part of the bubble.
For real values of $p_0$, above the threshold the real part of $b_T(p_0)$
can be obtained by taking the principal value in the right hand side of Eq. 
(\ref{Eq:buble_T}):
\begin{equation}
\textrm{Re}~b^>_T(p_0)=\frac{1}{4\pi^2} {\mathcal P}\int_{m_G(T)/T}^\infty dx
\frac{\sqrt{x^2-m_G^2(T)/T^2}}{p_0^2/4T^2-x^2}\frac{1}{\exp(x)-1}.
\label{rebT}
\end{equation}

Introducing the parametrization $p_0=\textrm{Re}p_0+i\varepsilon$ 
we can evaluate the imaginary part of $b_T(p_0)$ in the neighborhood of
 the real axis both for $\varepsilon>0$ (the physical prescription) and for
 $\varepsilon<0$:
\bea
&&\textrm{Im}\int_{m_G(T)}^\infty\frac{dq_0}{q_0}\sqrt{q_0^2-m_G^2(T)}
\frac{n(q_0)}{2q_0\pm (\textrm{Re}p_0+i\varepsilon)}\nonumber\\
&=&
-\pi\int_{m_G(T)}^\infty\frac{dq_0}{q_0} \sqrt{q_0^2-m_G^2(T)}n(q_0)
\delta(2q_0-|\textrm{Re}p_0|)
\left(\Theta(\pm\varepsilon)-\Theta(\mp\varepsilon)
\right).
\eea
With this one obtains:
\be
\textrm{Im}b_T(p_0)=-\frac{\textrm{sgn}(\varepsilon)}{8\pi}
\frac{\sqrt{(\textrm{Re}p_0)^2-4m_G^2(T)}}{\textrm{Re}p_0}n
\left(|\textrm{Re}p_0|/2\right)
\Big[\Theta\left(\textrm{Re}p_0-2m_G(T)\right)+
\Theta\left(-\textrm{Re}p_0-2m_G(T)\right)
\Big].
\label{Eq:imbT}
\ee

In order to ensure the continuity of the imaginary  parts
 an extra term has to be added to the expression used in the upper 
half-plane for the bubble:
\begin{equation}
b^<_T(p_0)=b^>_T(p_0)-\frac{i}{4\pi}n(p_0/2)\sqrt{1-\frac{4m_G^2(T)}{p_0^2}}.
\end{equation}

For later use (among others for the computation of the spectral function)
it is useful to write explicit expressions for the physical values of $b(p_0)$
 on the real positive axis. We have 
$b^>_T(p_0)=\textrm{Re}~b^>_T(p_0)+i\textrm{Im}~b^>_T(p_0)$, where the first 
term is given by Eq.(\ref{rebT}), and
\begin{equation}
\label{Eq:bTrealpart>0}
\textrm{Im}~b^>_T(p_0)=-\frac{1}{8\pi}
\sqrt{1-\frac{4m_G^2(T)}{p^2_0}}
n(p_0/2)\Theta\left(p_0-2m_G(T)\right).
\end{equation} 
The bubble is fully real below the threshold and it is clear that
the integral in Eq. (\ref{rebT}) is not 
singular for $p_0<2m_G(T)$. Its integrand has for $p_0=2m_G(T)$ an integrable 
square  root singularity: 
$$\displaystyle \textrm{Re}~b^>_T(p_0=2m_G(T))=
-\frac{1}{4\pi^2}\int_{m_G(T)/T}^\infty dx \frac{1}{\sqrt{x^2-m_G^2(T)/T^2}} 
\frac{1}{\exp(x)-1}.$$ 

For complex $p_0$ also in the lower halfplane one can use for $b_0^<(p_0)$
the expression written on the second line of Eq. (\ref{Eq:b0>R}) and the 
complete expression of $b^<_T(p_0)$ reads as
\begin{equation}
b^<_T(p_0)=\frac{1}{4\pi^2}\int_{m_G(T)/T}^\infty dx
\frac{\sqrt{x^2-m_G^2(T)/T^2}}{p_0^2/4T^2-x^2}\frac{1}{\exp(x)-1}-
\frac{i}{4\pi}n(p_0/2)\sqrt{1-\frac{4m_G^2(T)}{p_0^2}}.
\end{equation}

\section {The temperature dependence of the $\sigma$-pole}

In this section the temperature driven variation of the pole of
$G_\sigma(p_0)$, located in the fourth quadrant of the complex $p_0$ plane
 will be found. We shall 
argue that the generic scenario of its variation is realized independently
of what thermodynamical quantity would drive this variation. As an 
illustration of this we shall discuss the pole trajectory under the change
of the baryonic charge density implemented following Hatsuda,
Kunihiro,  and 
Shimizu \cite{hatsuda99}.

For the solution of the equation 
\begin{equation}
G_\sigma^{-1}(p_0)\left(1-\frac{\lambda_R}{6} b^<(p_0)\right)=
\left(p^2_0-m_G^2(T)\right)
\left(1-\frac{\lambda_R}{6}\Bigl(b_0^<(p_0)+b_T^<(p_0)\Bigr)\right)-
\frac{\lambda_R}{3}\Phi^2(T)=0
\label{sigma-pole}
\end{equation}
one obviously should know $\Phi(T)$ (and $m_G^2(T)=h/\Phi (T)$),  
calculable from the renormalized equation of state (\ref{Eq:ren_EoS}). 
This equation requires two inputs. The phenomenological (physical) input is 
$\Phi_0^2/m_{G0}^2=\Phi_0^3/h\equiv f_\pi^2/4m_{G0}^2\sim 0.11$, 
but in the equation of state also the normalization scale appears explicitly.

In order to simplify the formulae we have decided to choose the
absolute value of the pole location at $T=0$ for the normalization scale $M_0$.
Therefore one has to find first the $\sigma$-pole for this temperature and 
only then one can turn to the discussion of the finite temperature variation. 
 
\subsection{The poles of $G_\sigma(p_0)$ at $T=0$}

We parametrize the solution of the equation
\begin{equation}
G_\sigma^{-1}(p_0)\left(1-\frac{\lambda_R}{6} b_0^<(p_0)\right)=
\left(p_0^2-m_{G0}^2\right)\left(1-\frac{\lambda_R}{6}b_0^<(p_0)\right)-
\frac{\lambda_R}{3}\Phi_0^2=0
\label{Eq:zeroT}
\end{equation}
in the form $p_0=M_0\exp(-i\varphi_0)$, $0<\varphi_0<\pi/2$. 
Note, that the renormalization scale is fixed in proportion to $m_{G0}$ once a
 renormalized coupling is chosen.

One can introduce instead of $M_0$ and $\varphi_0$ a more convenient
parametrization $p_0=2m_{G0}+\bar{M}_0\exp(-i\bar\varphi_0)$, where 
$\bar M_0$ and $\bar \varphi_0$ are uniquely determined by 
$M_0$ and $\varphi_0$. Attention has to be payed when switching from one 
parametrization to the other because changing the scale $M_0$ means changing 
the renormalized coupling constant $\lambda_R$.

In addition one has to deal with care when realizing numerically
the analytic continuation. For the square root
and for the evaluation of the argument of the complex numbers appearing in 
$b_0^<(p)$
one has to choose a phase convention which ensures the continuous variation 
of the complex phase of the final complex number with $\varphi_0$.
Good guidance for how to define the
argument of complex numbers in the process of evaluating complicated
multivalued functions is provided by their series expansion near the
positive real axis above the threshold.

\begin{figure}[htbp]
\begin{center}
\includegraphics[width=9cm]{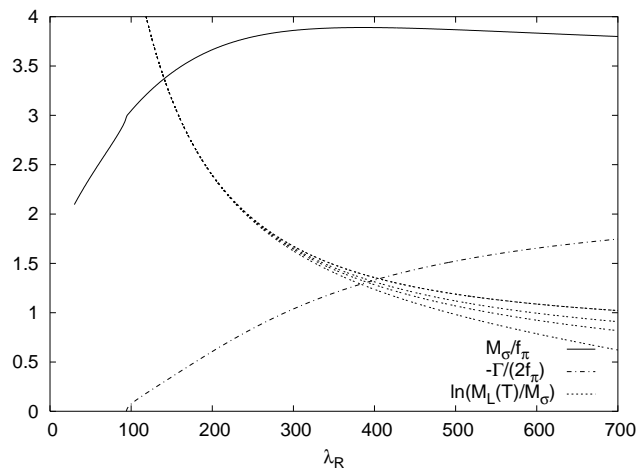}
\end{center}
\vspace*{-0.5cm}
\caption{\at{The real and imaginary parts of the physical poles at $T=0$. 
Also shown is the logarithm of the
tachyon pole position in proportion to the mass of $\sigma$ for various
temperatures. The lines appear in the same  order downward from above
on the right side of the figure as the labels in the key.}}
\label{Fig:T0-poles_h}
\end{figure}

The output is $M_0/\Phi_0$ and $\varphi_0$ in terms of which we 
obtain the mass $M_\sigma=M_0\cos\varphi_0$ and the width 
$\Gamma=2M_0\sin\varphi_0$ of the $\sigma$ as the real and imaginary 
parts of the pole. These are shown as a 
function of the renormalized coupling $\lambda_R$  in 
Fig.\ref{Fig:T0-poles_h}. At $\lambda_R=400$ one finds the ratio 
$M_\sigma/\Gamma\sim 1.4$ with $M_\sigma=3.95f_\pi$. These values are away 
from the phenomenological expectations \at{\cite{close02,tornqvist02}}, 
but these are the best values we can 
reach in the leading large $N$ approximation. The use of higher values of 
$\lambda_R$ which appear to be closer to the observed numbers might not be 
advisable since the a tachyonic pole described below comes very close to the 
scale $M_0$ for that coupling region. We will use $\lambda_R=400$ henceforth 
in the finite temperature calculations.

Scalar theories are known to have a tachyonic pole related to the
Landau-ghost \cite{boyan99}, that is a zero of
the inverse propagator on the positive imaginary axis $p_0=iM_L$.
Equation (\ref{Eq:zeroT}) takes the following form when looking for $M_L$:
\begin{equation}
\big(M_L^2+m_{G0}^2\big)\left(1-\frac{\lambda_R}{6}b_0^{>}(iM_L)\right)+
\frac{\lambda_R}{3}\Phi_0^2=0.
\label{tachyon}
\end{equation}
with a fully real expression for $b_0^>(iM_L)$:
\begin{equation}
\displaystyle
b_0^>(iM_L)=\frac{1}{16\pi^2}\left[\ln\frac{m_{G0}^2}{M_0^2}-
\sqrt{1+\frac{4m_{G0}^2}{M_L^2}}
\ln\frac{\sqrt{1+4m_{G0}^2/M_L^2}-1}{\sqrt{1+4m_{G0}^2/M_L^2}+1}\right].
\end{equation}

With a suitable parametrization, for example, $M_L/M_0=\exp(z)$, one can
solve Eq.(\ref{tachyon}) for $z$ at a given value of $\Phi_0^2/m_{G0}^2$
and using the value of $M_0/\Phi_0$ obtained by solving Eq.(\ref{Eq:zeroT}).
The logarithm of the ratio $M_L/M_\sigma$ whose value restricts the 
range of validity of the theory is shown in Fig.\ref{Fig:T0-poles_h} as a 
function of $\lambda_R$, not only for $T=0$ but also for some non-zero 
temperatures.

\subsection{Finite $T/n_B$ behavior of the $\sigma$-pole}

The features of the numerical solution of Eq.(\ref{sigma-pole}) will be 
discussed in the main part of the present subsection. 
It will be pointed out that when moving on the second Riemann sheet its root 
approaches and eventually hits at a certain temperature the unphysical 
real axis {\it below} the two-pion threshold. For the
determination of the poles on this piece of the real
axis one can evaluate the analytic functions directly. The real
solution provides
a useful check of the solution based on the complex equation. We give here
the corresponding formulae explicitly.

In both parametrizations of the pole one has on the real axis below
the threshold ($p_0<2m_G(T)$), $\varphi=\pi, \bar\varphi=\pi$, respectively.
 Then one has $(1-4m_G^2(T)/p_0^2)^{1/2}=-i(4m_G^2(T)/p_0^2-1)^{1/2}\equiv
-iQ$. Using this in Eq. (\ref{Eq:b0>R}) one obtains:
\bea
\nonumber
-\sqrt{1-\frac{4m_G^2(T)}{p_0^2}}
\left(\ln\frac{1-\sqrt{1-4m_G^2(T)/p_0^2}}{1+\sqrt{1-4m_G^2(T)/p_0^2}}+
i\pi\right)
=iQ\left(\ln\frac{Q^{-1}+i}{Q^{-1}-i}+ i\pi\right)&&\\
=-2Q\textrm{arccot}(Q^{-1})-\pi Q=
2Q(\textrm{arccot}(Q)+\pi).&&
\eea
So, on the real axis of the second Riemann sheet, below
the threshold one has:
\bea
b_0^{<-}(p_0)&=&\frac{1}{16\pi^2}\left[\ln\frac{m_G^2(T)}{M_0^2}+
2\sqrt{\frac{4m_G^2(T)}{p_0^2}-1}\left(
\textrm{arctan}\left(\frac{4m_G^2(T)}{p_0^2}-1\right)^{-\frac{1}{2}}-\pi\right)
\right],\\\displaystyle
b_T^{<-}(p_0)&=&-\frac{1}{4\pi}\sqrt{\frac{4m_G^2(T)}{p_0^2}-1}\frac{1}
{\exp(p_0/2T)-1}+\frac{1}{4\pi^2}\int_{m_G(T)/T}^\infty dx
\frac{\sqrt{x^2-m_G^2(T)/T^2}}{(p_0^2/4T^2-x^2)}\frac{1}{\exp(x)-1}.
\eea  
(Here we use a somewhat redundant superscript ``$^-$'' on $b^<(p)$ in order 
to emphasize that the formulae refer to a continuation below the threshold).

The evaluation of $b^>(p_0)$ below threshold on the real axis becomes relevant
for temperatures when the pole ``climbs up'' from the second onto the first
Riemann sheet and represents a stable $\sigma$ particle.
When approaching this portion of the real axis from the physical upper 
halfplane one trivially finds the corresponding real expression.
For real $p_0$ the equations which determine the poles on the physical and 
unphysical real axes below the threshold ($p_0<2m_G(T)$) are the following:
\begin{equation}
\left(p^2_0-m_G^2(T)\right)
\left(1-\frac{\lambda_R}{6}\Bigl(b_0^>(p_0)+b_T^>(p_0)\Bigr)\right)-
\frac{\lambda_R}{3}\Phi^2(T)=0
\label{real-above}
\end{equation}
and
\begin{equation}
\left(p^2_0-m_G^2(T)\right)
\left(1-\frac{\lambda_R}{6}\Bigl(b_0^{<-}(p_0)+b_T^{<-}(p_0)\Bigr)\right)-
\frac{\lambda_R}{3}\Phi^2(T)=0,
\label{real-below}
\end{equation}
respectively.

\begin{figure}[htbp]
\begin{center}
\includegraphics[width=9cm]{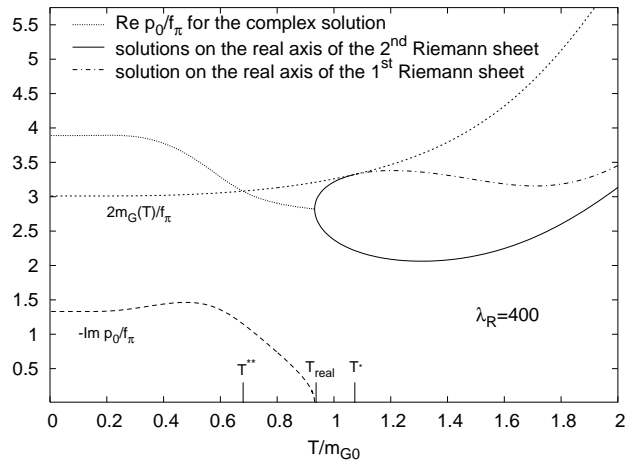}
\end{center}
\vspace*{-0.7cm}
\caption{The temperature dependence of the real and imaginary parts of
  the $\sigma$-pole}
\label{Fig:T-poleR}
\end{figure}

In Fig.\ref{Fig:T-poleR} the trajectory of the real part of the pole 
position is shown as it evolves with the temperature. In the same figure also 
the $T$-dependent location of the two-pion threshold appears. One notes 
that $\textrm{Re}(p_0)$ crosses below the actual position of the threshold for
$T^{**}/m_{G0}\sim 0.68$. As it is seen from Fig.\ref{Fig:T-poleR} 
$\textrm{Im}(p_0)$ hardly diminishes until this temperature is reached, 
therefore the relative broadening of $\sigma$ actually increases.
Similar conclusions where drawn in the context of chiral perturbation
theory in \cite{Estrada}. Above $T^{**}$ the imaginary part decreases
faster and the pole position is landing on the real  axis for 
$T_{real}/m_{G0}\sim 0.93$.

One can study the solutions of Eq.(\ref{sigma-pole}) also in the upper $p_0$ 
halfplane on the {\it second} Riemann sheet. One finds that starting from $T=0$,
there exists a ``mirror''-root, which arrives at the same point of the
real axis for $T=T_{real}$. This collision of the poles results in two
oppositely moving real solutions for higher temperatures. The solutions of
Eq.(\ref{real-below}) fully confirm this scenario.

The pole moving upwards catches up with the threshold for $T^*/m_{G0}=1.074$.
(The other pole first moves downwards, later its motion changes direction, but
it lags behind the position of the threshold in the whole temperature range 
of interest, \at{see. Fig. \ref{Fig:T-poleR}}.) It does not stop there, but moves further with increasing
temperature, now on the real axis of the physical Riemann sheet. 
This part of the scenario is 
confirmed also by the direct solution of Eqs.(\ref{real-below}) and
(\ref{real-above}). The 
$T$-dependent position of the stable physical $\sigma$ particle is also 
displayed in Fig.\ref{Fig:T-poleR}. One has to notice that all scales
increase with the temperature, therefore the tachyon pole puts a strict
temperature limit to the validity of the proposed effective treatment of the 
pion-sigma system.

\begin{figure}[htbp]
\begin{center}
\includegraphics[width=9cm]{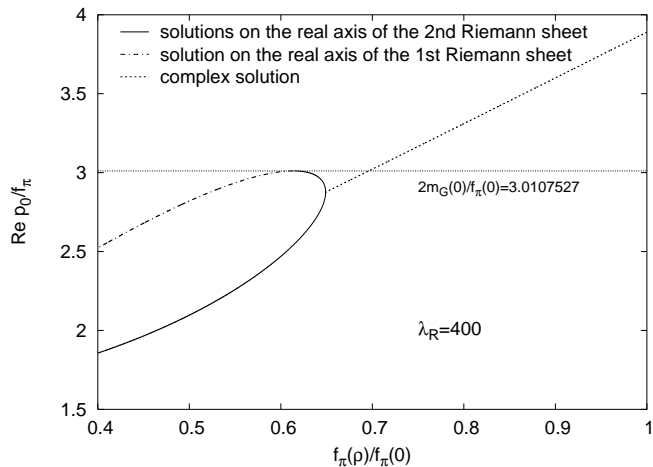}
\end{center}
\vspace*{-0.7cm}
\caption{The dependence of the real part of the pole position on the variation
 of $f_\pi$ due to non-zero baryonic density}
\label{Fig:rho-poleR}
\end{figure}

The scenario obtained is clearly different from the one suggested by 
Hatsuda, Kunihiro, and their collaborators \cite{hatsuda01}.
 In their various approximate descriptions the real part of the pole position
never goes below the two-pion threshold until it has a finite imaginary 
part. In order to test the generic nature of the pole evolution found
above, we have followed the procedure proposed by Hatsuda {\it et al.}
\cite{hatsuda99} for the
introduction of finite baryonic charge density $n_B$ into the effective 
pion-sigma dynamics. In summary, the nonzero value of $n_B$ results in a
rescaling of the vacuum expectation value $\Phi$ \cite{brockmann96}. 
At $T=0$, for low densities they propose
\begin{equation}
\Phi (n_B)=(1-Cn_B)\Phi_0,
\end{equation}
with $C=0.2-0.3$. With the parametrization $p_0=\eta M_0\exp(-i\varphi)$ 
we have solved Eq.(\ref{Eq:zeroT}) for $\eta$ and $\varphi$  with $\Phi (n_B)$ 
replacing $\Phi_0$ everywhere. This equation is much simpler, since the 
details of its continuation onto the second Riemann sheet are self-evident.
Still, we find qualitatively the same pattern for the pole trajectory as one
can see in Fig.\ref{Fig:rho-poleR}. In another test the strength of the
explicit symmetry breaking ($h$) was gradually decreased. The distance of the
point where the complex solution arrives to the real axis from the
two-pion threshold monotonically
decreases with the decrease of $h$. In the chiral limit it approaches 
smoothly the origin as it should for a true phase transition. It is worthwhile 
to point out that the most recent dispersive investigation of 
Yokokawa {\it et al.}\cite{yokokawa02} investigates exactly this limit.
They find the same smooth behavior, therefore there is no conflict
between the results of the two approaches yet. It will be interesting
to see the effect of explicit symmetry breaking in their approach.

\begin{figure}[htbp]
\begin{center}
\includegraphics[width=9cm]{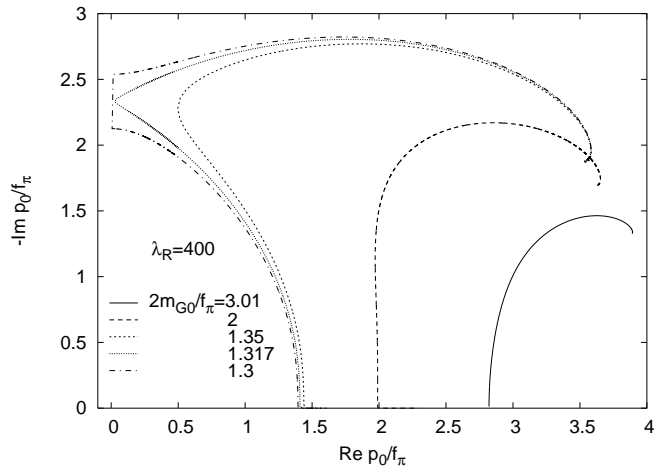}
\end{center}
\vspace*{-0.7cm}
\caption{Trajectory of the complex $\sigma$-pole for various values of
$2m_{G0}/f_\pi$. \at{Note the tendency of the trajectory to
approach closer to the imaginary axis as $m_{G0}$ decreases.}}
\label{TmG0-less}
\end{figure}

Finally, we have studied systematically the deformation of the $T$-driven 
pole trajectory when $2m_{G0}/f_\pi$ is decreased gradually with $f_\pi$ kept
constant. A quite interesting pattern appears in Fig.\ref{TmG0-less}. The 
trajectory reaches closer to the negative imaginary axis as the strength of 
the explicit symmetry breaking diminishes, before it turns to the real
axis, and eventually ends at the two-pion threshold. At some value
$1.317>2m_{G0,imag}/f_\pi>1.316$ it touches first this axis, 
but is ``reflected'' from it back into the fourth quarter.
For higher temperatures it will have again a non-zero real part.

In order to understand what happens, it is convenient to search directly
for poles on the negative imaginary axis. It turns out that already for
$2m_{G0}/f_\pi =3.01$ one has an infinite number of approximately
equidistantly located poles along this axis  for finite temperature. 
Below we call the pole located the closest to the origin the $\sigma^*$-pole. 
Already at low temperature the distance of all these poles from the threshold 
is much larger, than that of the $\sigma$-pole. This explains why the latter 
dominates near $T=0$ the behavior of the spectral function, as will be argued 
in the next section. The distance further increases with the increase of the 
temperature.

It turns out that for the above quoted value of $2m_{G0,imag}/f_\pi$ the 
$\sigma$-pole and its mirror from the third quarter touch the imaginary axis 
exactly at the location of the highest (negative) imaginary $\sigma^*$-pole at 
that  temperature. The result of the {\bf ``$\sigma -\sigma^*$''} collision 
is the reflection of the two complex poles back into their respective quarters.

Further decreasing $2m_{G0}/f_\pi$ the pair of mirror poles arrives
onto the negative imaginary axis below the highest imaginary pole. 
\at{(In Fig. \ref{TmG0-less} it is above $\sigma^*$, since the negative 
imaginary axis is directed upwards.)}
The 
colliding poles now give rise to a purely imaginary pair, one member of which
moves towards the origin the other one moving the opposite direction.
The pole moving towards the origin collides at some higher temperature
with the oppositely moving genuinely imaginary $\sigma^*$-pole, and
they are pushed back into the complex quarters as mirror poles. Eventually, 
the pole in the fourth quarter will land on the real axis and moves up to the
threshold for  $T^*(m_{G0})$, where it is converted into a stable particle 
pole on the physical sheet, as described earlier in this section.
In view of the multiple pole collisions it is clear that the stable high-$T$ 
$\sigma$ particle is not directly related to the $T=0$ complex $\sigma$-pole.

In the chiral limit the $\sigma$-pole reaches the origin and describes
the  phase transition restoring the chiral symmetry of the model. This simple
trajectory has been already discussed in our previous publication
\cite{patkos02}. The presence of explicit symmetry breaking with
realistic strength led to a rather spectacular change in this scenario. 
It is a valid question whether a smooth continuous deformation of the
pole trajectory connects the case of the explicit symmetry breaking
with the chiral limit.

In the region, when the pion mass is much smaller than the temperature for
which the complex pole becomes purely imaginary, one can find analytically
the first $\sigma^*$-pole in the infinite sequence described above. Its location is
given approximately as $p_{\sigma^*}^2\sim m_{G}^3(T)T/\Phi^2(T)$.
In view of the fact that $T/\Phi (T)\sim{\cal O}(1)$, this pole goes faster 
to zero than the pion mass.  Our numerical study shows that the point where 
the pole lands on the negative imaginary axis does not change more than 10\% 
between the pion mass $m_{G0,imag}$ and the chiral limit. Therefore the
interval of temperatures for which the pole moves on the imaginary
axis increases with decreasing pion mass (see Fig.\ref{TmG0-less}).
Eventually for $m_{G0}=0$ the highest negative imaginary pole stays
(with zero residuum) in the origin and the scenario characterizing the
chiral limit sets in smoothly.

For symmetry breaking much smaller than the parameter characterizing the 
onset of the dynamical scaling in the chirally symmetric case one can even 
experience the realization of the scaling behavior.

To some extent the above complicated trajectory is to be expected, since it 
is highly ``improbable'' that the roots of a complex equation would move
smoothly to a specific point (e.g., the two-pion threshold) of the real axis, 
irrespective of the variation of its parameters. The limitations of the true 
resonance interpretation of the $\sigma$-pole will be  discussed in the next 
section, when the correlation of its location with the measurable spectral 
function $\rho_\sigma$ will be discussed. 

\section{The $T$ dependence of the spectral function $\rho_\sigma$}

The spectral function of the order parameter field $\sigma$ is defined
using the expression of the propagator in the physical half-plane as:
\begin{equation}
\rho_\sigma (p_0,{\bf p},T)=-{1\over\pi}
\lim_{\varepsilon\rightarrow +0}{\rm Im}G_\sigma(p_0+i\varepsilon,{\bf p},T).
\end{equation}

The leading order large $N$ expression of spectral function, at $\p=0$, 
is given by
\bea
&
\rho_\sigma(p_0,0,T)=\lambda_R^2\Phi^2(T)\textrm{Im}b^>(p_0)/18\pi\times
\nonumber\\
&\Bigl [\left[(p_0^2-m_G^2(T))\left(1-\frac{\lambda_R}{6}
\textrm{Re}b^>(p_0)\right)-\frac{\lambda_R}{3}\Phi^2(T) 
\right]^2+\left(p_0^2-m_G^2(T)\right)^2\frac{\lambda_R^2}{36}
(\textrm{Im}b^>(p_0))^2\Bigr ]^{-1}.
\label{spectral}
\eea

We have seen that $\textrm{Im}b^>(p_0)\neq0$ only for $p_0>2m_G(T)$, if there 
is no stable particle pole below the threshold. Therefore the
spectral function is nonzero only for $p_0$ values above the threshold until
$T<T^*$. For this reason the numerator and the second term in the denominator 
of Eq.(\ref{spectral}), that is $\textrm{Im}b^>(p_0)$ goes to zero when 
$p_0\rightarrow 2m_G(T)+0.$ 

\begin{figure}[htbp]
\begin{center}
\includegraphics[width=9cm]{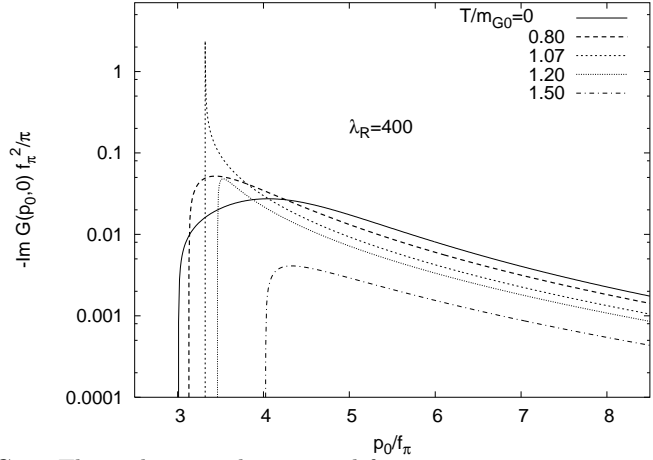}
\end{center}
\vspace*{-0.7cm}
\caption{The scalar-isoscalar spectral function at various temperatures}
\label{Fig:rho}
\end{figure}

The scalar-isoscalar spectral function is displayed in Fig.\ref{Fig:rho}.
One notices the shift of its maximum towards the two-pion threshold with 
increasing temperature, though its width does not decrease initially. For 
$T>T^{**}$ the shape of $\rho_\sigma$ becomes cuspier. Finally, a very high value 
of the maximum is experienced numerically around $T/m_{G0}\sim 1.07$. Above 
this temperature the value of the maximum gradually diminishes and its 
location shifts increasingly farther from the threshold towards larger $p_0$.

For a qualitative interpretation of $\rho_\sigma$ one verifies that for the 
temperature $T^*\approx1.074m_{G0}$, the first term of the denominator 
of Eq.(\ref{spectral}) vanishes
at the threshold $p_0=2m_G(T^*)$, that is Eq.(\ref{real-above}) is equivalent
to the condition for the vanishing of the first term in the denominator of
Eq. (\ref{spectral}):
\begin{equation}
\frac{6}{\lambda_R}-b^>_0(2m_G(T^*))-b^>_T(2m_G(T^*))-\frac{2}{3}
\frac{\Phi^2(T^*)}{m_G^2(T^*)}=0.
\end{equation}

Because the term containing the real part of the bubble vanishes more
rapidly at $T^*$ as $p_0\rightarrow 2m_G(T^*)$ than the term containing
the imaginary part of the bubble, the behavior of the spectral function 
around the threshold is dominated by the imaginary part of the bubble
$\rho_\sigma(p_0,0,T^*)\sim 1/\textrm{Im}b^>(p_0)=
1/(1-4m_G^2(T^*)/p_0^2)^{1/2}.$ This is formally the same behavior exploited 
by Hatsuda {\it et al.} \cite{hatsuda99}
when arguing in favor of the generic nature of the
threshold enhancement phenomenon. It is worthwhile to emphasize, however, the 
obvious fact that according to our
calculation the spectral function has nothing to do with the imaginary
part of the second sheet pole 
which is a purely real (unphysical) quantity in this temperature
regime, and the singular behavior is due to the fact that the pole
along the unphysical real axis moves towards the threshold when
$T\rightarrow T^*$.

One might suspect, that the coincidence
 of the pole position at $T=T^*$ with the threshold would lead to a stronger
$\sim \delta (p_0-2m_G(T^*))$ singularity for this temperature. This is
not true since one can easily demonstrate the vanishing of its
residuum for $T=T^*$. The residuum of the stable $\sigma$ pole 
appearing on the physical sheet for $T>T^*$ continuously increases with the 
temperature.

The near threshold enhancement of the spectral function is maximal at $T^*$.
When $T>T^*$, the position of the maximum moves away from the threshold and 
its height diminishes. This is consistent with the requirement arising from the
sum rule, $\int dp_0p_0\rho_\sigma (p_0)=1$ in presence of a stable $\sigma$-pole
with increasing residuum. 

\begin{figure}[htbp]
\begin{center}
\includegraphics[width=9cm]{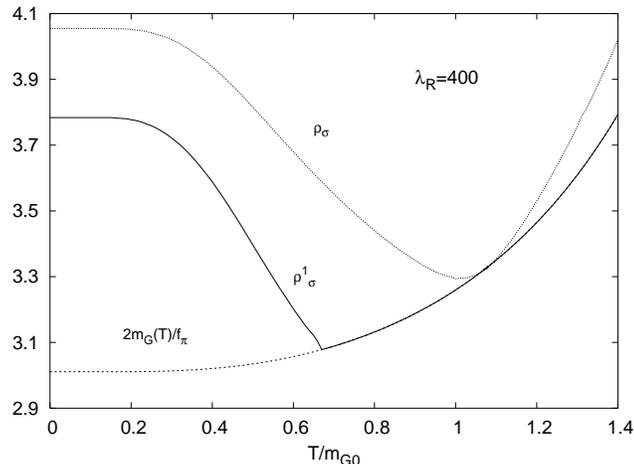}
\end{center}
\vspace*{-0.7cm}
\caption{The temperature dependence of the locations of the maxima of $\rho_\sigma$
and of $\rho^1_\sigma$ as compared to the two-pion threshold}
\label{Fig:T-maximum}
\end{figure}

The enhancement sets in gradually and for $T>T^{**}$ the
maximum of $\rho_\sigma$ stays very close to the actual threshold position.
We can argue rather convincingly for a certain physical significance of
$T^{**}$, when displaying $\rho_\sigma^1(p_0)\equiv\rho_\sigma(p_0)/
\sqrt{1-4m_G^2(T)/p_0^2}$. The argument for this operation is the fact that
the phase space volume is just proportional to the factor divided out.
In Fig.\ref{Fig:T-maximum} one can follow the
position of the maxima of $\rho_\sigma$ and $\rho_\sigma^1$ relative to the two-pion 
threshold as a function of the temperature. One sees that the position of
${\rm max}(\rho_\sigma)$ touches the threshold only in a single point $T=T^*$. On 
the other hand the position of ${\rm max}(\rho_\sigma^1(T))$ approaches $2m_G(T)$ 
rather steeply, but starting from $T=T^{**}$ the distance is found numerically
always smaller than  $10^{-5}$.

In conclusion of this section one sees that the qualitative changes in the 
spectral function can be well interpreted with the help of the scalar-isoscalar
pole located in the lower $p_0$ halfplane. The threshold enhancement
occurs in an extended temperature region, $T^{**}<T<T^*$.
Any analysis attempting  the reconstruction of the in-medium
$\sigma$ resonance
from some experiment is necessarily based on the behavior of the 
scalar-isoscalar spectral function. From the above discussion it is suggestive
 that by the enhanced signal coming from the neighborhood of the threshold
one would be led to the conclusion that in the temperature range 
$T^{**}<T<T^*$ the ``$\sigma$'' moves together with the two-pion threshold.  
\at{But no trace of any Lorentzian resonance shape can be detected in the
spectral function.}

\section{Conclusions}

In this paper we have presented all the technical details of analyzing
the pole trajectory of the propagator describing the fluctuations of the
chiral order parameter in the linear sigma model with the help of a
leading large $N$ approximation. It is worth noting that the propagator of the
composite field $\left(\phi^a({\bf x},t)\right)^2$ has the same poles
as that  of $\phi$ in
the broken symmetry phase (see Ref.\cite{patkos02} and references therein).
Detailed mapping of the trajectory under the
variation of the temperature/baryonic density as well as the parameter
controlling the
explicit breaking of the chiral symmetry leads us to conjecture that
generically the real part of the complex pole will decrease below the 
two-pion threshold energy with a finite imaginary part. The application of
higher order corrections in the large $N$ expansion are expected to 
give no more than 25\% correction. Therefore only quantitative change
may happen in the scenario presented in our paper.

Our main result is the prediction that in a whole
temperature range $T^{**}<T<T^*$ the maximum of the spectral function
is rather close to the actual location of the two-pion threshold. In
this sense one can speak about an interval of threshold enhancement.
The point of maximal enhancement is $T=T^*$.

The usual connection between the second Riemann sheet pole and the
spectral function is restricted to the temperature range
$T<T^{**}$. In this region the spectral
function has an approximately Lorentzian shape, its maximum is centered
nearly at the location of the real part of the pole, its width scales
with its imaginary part.  When 
approaching $T^{**}$ from below, the approximately Lorentzian shape of
the spectral function will be distorted, in particular it is loosing
its symmetry. With considerable compromise one can extend the
$\sigma$-particle
interpretation of the second Riemann sheet pole up to $T^{**}$.

The complete loss of this characteristic means that no resonance
interpretation can be given to the pole in the $T^{**}<T<T^*$
range. On the other hand any phenomenological analysis of a particle signal 
proportional to the spectral function will be peaked near the two-pion
threshold in the above temperature interval. The real-time propagator in
this channel will be dominated by the contribution of frequencies
slightly above $2m_G(T)$. In this sense an increasingly narrow
 $\sigma$-signal can be detected at $\omega\approx 2m_G(T)$
in the final two-pion spectra.

 Above $T^*$ it is a stable physical $\sigma$-particle, which
 reappears. Accordingly, the peak of the spectral function
 diminishes. By the increased effect of the phase space factor its
 maximum will be reached again at frequencies farther above the
 two-pion threshold.

\section*{Acknowledgments}

This research has been supported by the research contract
OTKA-T037689 of the Hungarian Research Fund.

\end{document}